\documentclass[aps,prl,twocolumn,superscriptaddress,groupedaddress]{revtex4}  % for review and submission
\usepackage{graphicx}  % needed for figures
\usepackage{dcolumn}   % needed for some tables
\usepackage{bm}        % for math
\usepackage{amssymb}   % for math
\usepackage{color}
\usepackage{braket}
\usepackage{pdfpages}
\usepackage{pgffor}

\newcommand{\dd}{\mathrm{d}}

\begin{document}

\title{Dynamics of a qubit while simultaneously monitoring its relaxation and dephasing}

\author{Q. Ficheux}

\affiliation{Universit\'e Lyon, ENS de Lyon, Universit\'e Claude Bernard, CNRS, Laboratoire de Physique, F-69342
Lyon, France}
\affiliation{Laboratoire Pierre Aigrain, D\'epartement de physique de l'ENS, \'Ecole normale sup\'erieure, PSL Research University, Universit\'e Paris Diderot, Sorbonne Paris Cit\'e, Sorbonne Universit\'es, UPMC Univ. Paris 06, CNRS,
75005 Paris, France
}

\author{S. Jezouin}
\affiliation{Laboratoire Pierre Aigrain, D\'epartement de physique de l'ENS, \'Ecole normale sup\'erieure, PSL Research University, Universit\'e Paris Diderot, Sorbonne Paris Cit\'e, Sorbonne Universit\'es, UPMC Univ. Paris 06, CNRS,
75005 Paris, France
}

\author{Z. Leghtas}

\affiliation{Centre Automatique et Syst\`emes, Mines ParisTech, PSL Research University,
60 Boulevard Saint-Michel, 75272 Paris Cedex 6, France.}
\affiliation{Laboratoire Pierre Aigrain, D\'epartement de physique de l'ENS, \'Ecole normale sup\'erieure, PSL Research University, Universit\'e Paris Diderot, Sorbonne Paris Cit\'e, Sorbonne Universit\'es, UPMC Univ. Paris 06, CNRS,
75005 Paris, France
}
\affiliation{QUANTIC team, INRIA de Paris, 2 Rue Simone Iff, 75012 Paris, France}

\author{B. Huard}
\email{benjamin.huard@ens-lyon.fr}
\affiliation{Universit\'e Lyon, ENS de Lyon, Universit\'e Claude Bernard, CNRS, Laboratoire de Physique, F-69342
Lyon, France}
\affiliation{Laboratoire Pierre Aigrain, D\'epartement de physique de l'ENS, \'Ecole normale sup\'erieure, PSL Research University, Universit\'e Paris Diderot, Sorbonne Paris Cit\'e, Sorbonne Universit\'es, UPMC Univ. Paris 06, CNRS,
75005 Paris, France
}

\date{\today}

\begin{abstract}
{\large \textbf{Abstract}}

Decoherence originates from the leakage of quantum information into external degrees of freedom. For a qubit the two main decoherence channels are relaxation and dephasing. Here, we report an experiment on a superconducting qubit where we retrieve part of the lost information in both of these channels. We demonstrate that raw averaging the corresponding measurement records provides a full quantum tomography of the qubit state where all three components of the effective spin-1/2 are simultaneously measured. From single realizations of the experiment, it is possible to infer the quantum trajectories followed by the qubit state conditioned on relaxation and/or dephasing channels. The incompatibility between these quantum measurements of the qubit leads to observable consequences in the statistics of quantum states. The high level of controllability of superconducting circuits enables us to explore many regimes from the Zeno effect to underdamped Rabi oscillations depending on the relative strengths of driving, dephasing and relaxation.
\end{abstract}

\maketitle

{\large \textbf{Introduction}}

\begin{figure}[!ht]
\includegraphics[width=9cm]{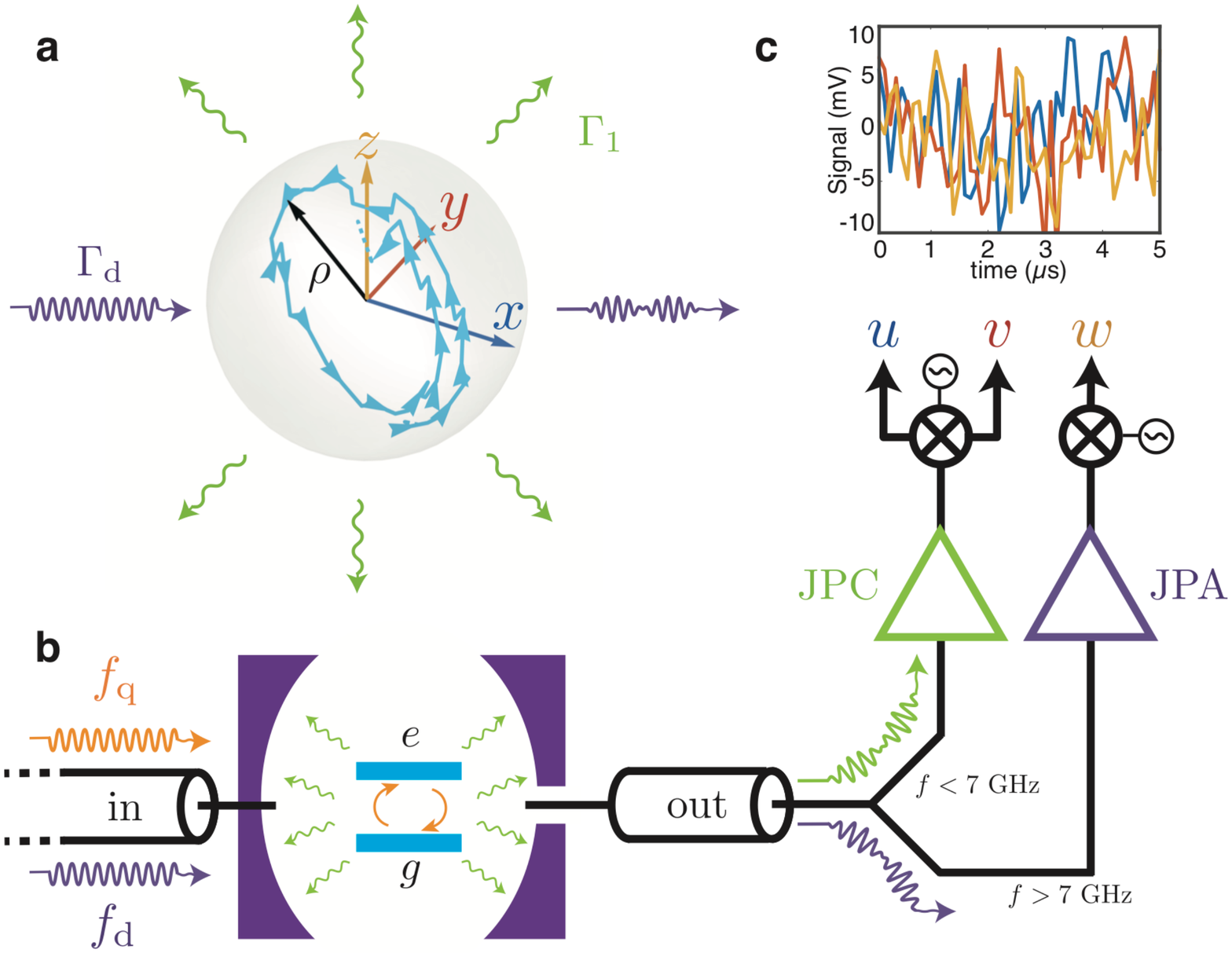}
\caption{Measurement setup and quantum trajectory resulting from its outputs. \textbf{a} Bloch vector representation of a qubit whose state is described by a density matrix $\rho_t=\left(\mathbf{1}+x(t)\sigma_x+y(t)\sigma_y+z(t)\sigma_z\right)/2$. A quantum trajectory $\rho_t$ is represented as a blue line. The qubit decoherence can be modeled as originating from a relaxation channel at a rate $\Gamma_1$ and a dispersive measurement channel at a rate $\Gamma_\mathrm{d}$. \textbf{b} A superconducting qubit in a cavity is driven by two microwave signals at the weakly coupled input. The one at qubit frequency $f_\mathrm{q}=5.353~\mathrm{GHz}$ (orange) induces Rabi oscillations of the qubit at frequency $\Omega$. The one at cavity frequency $f_\mathrm{d}=7.761~\mathrm{GHz}$ (purple) leads to a dispersive measurement of the qubit state along $\sigma_z$. A diplexer at the strongly coupled output port separates the outgoing signals depending on their frequency. The radiation at $f_\mathrm{q}$ that is spontaneously emitted by the qubit is processed by a Josephson Parametric Converter (JPC)~\cite{Bergeal,Roch2012} so that a following heterodyne measurement reveals the two quadratures $u(t)$ and $v(t)$ of the fluorescence field~\cite{PhysRevLett.112.180402,PhysRevX.6.011002,campagneibarcq:tel-01248789}. The transmitted signal at $f_\mathrm{d}$ is processed by a doubly pumped Josephson Parametric Amplifier (JPA)~\cite{Kamal2009a,Murch2013a} with a pump phase such that a following homodyne measurement reveals the quadrature $w(t)$ of the field at $f_\mathrm{d}$. \textbf{c} Measurement records $u$ (blue), $v$ (red) and $w$ (yellow) as a function of time for one realization of the experiment. These records feed the stochastic master equation (\ref{SME}), which leads to the trajectory in \textbf{a}\label{fig1}}
\end{figure}

Decoherence can be understood as the result of measurement of a system by its environment. For a qubit, the two main sources of decoherence are relaxation by spontaneous emission and dephasing that can be modeled by unmonitored readout of coupled quantum systems (Fig.~\ref{fig1}a). What becomes of the qubit state if, instead of disregarding the information leaking to the environment, we continuously monitor both decoherence channels? Owing to measurement backaction, the knowledge of the measurement record then leads to a stochastic quantum trajectory of the qubit state for each single realization of an experiment~\cite{Carmichael1993,Barchielli2009,Wiseman2009}. Recently, diffusive quantum trajectories were observed following the continuous homodyne or heterodyne measurements of either a dephasing channel~\cite{Murch2013a,Hatridge2013,DeLange2014,Weber2014,Weber2016,Chantasri2016a} or a relaxation channel~\cite{PhysRevX.6.011002,Naghiloo2016}.

Here we report an experiment in which we have simultaneously monitored the spontaneous emission of a superconducting qubit by heterodyne measurement (relaxation channel) and the transmitted field through a dispersively coupled cavity by homodyne measurement (dephasing channel). We demonstrate that the average outcomes of these two non-projective measurements are the three coordinates $x$, $y$ and $z$ of the Bloch vector. It is remarkable that a full quantum tomography can be obtained at any time by simply raw averaging measurement outcomes of many realizations of a single experiment despite the incompatibility of the three observables that characterize a qubit state. For single realizations the resulting quantum trajectories show signatures of the incompatibility between the measurement channels, therefore extending the previously explored case of two incompatible measurement outcomes~\cite{PhysRevX.6.011002,Hacohen-Gourgy2016} to the case of three spin directions. By varying the drive amplitudes at the cavity and qubit transition frequencies, we are able to reach a variety of regimes corresponding to different configurations for $\Omega/\Gamma_1$ and $\Gamma_\mathrm{d}/\Gamma_1$, where $\Omega$ is the Rabi frequency, $\Gamma_1$ the fixed relaxation rate and $\Gamma_\mathrm{d}$ the dephasing rate. This work hence provides a textbook experimental demonstration of quantum measurement backaction on a qubit with incompatible and simultaneous measurements.

{\large \textbf{Results}}

\textbf{Description of the experiment.} Two parallel detection setups operate via spatially separated measurement lines (see Fig.~\ref{fig1}b). The fluorescence heterodyne detection setup enables the measurement of both quadratures $u(t)$ and $v(t)$ of the spontaneously emitted field out of a 3D transmon qubit~\cite{Paik2011a}. The  complex amplitude of the emitted field is on average proportional to the expectation of the qubit lowering operator $\sigma_-=(\sigma_x-i\sigma_y)/2$ so that $u$ and $v$ are on average proportional to the expectations of $\sigma_x$ and $\sigma_y$. Here, $\sigma_x=\ket{g}\bra{e}+\ket{e}\bra{g}$, $\sigma_y=i\ket{g}\bra{e}-i\ket{e}\bra{g}$ and $\sigma_z=\ket{e}\bra{e}-\ket{g}\bra{g}$ are the qubit Pauli operators, where $\ket{g}$ and $\ket{e}$ are the ground and excited states. For a single realization, the measurement outcomes read~\cite{PhysRevX.6.011002}
\begin{equation}
\left\{\begin{array}{rcl}
u(t)\dd t&=&\sqrt{\eta_\mathrm{f}\Gamma_1/2}x(t)\dd t+{\dd W_u(t)}\\
v(t)\dd t&=&\sqrt{\eta_\mathrm{f}\Gamma_1/2}y(t)\dd t+{\dd W_v(t)}
\end{array}
\right.,\label{uvrecords}
\end{equation}

where $\Gamma_1=(15~\mu\mathrm{s})^{-1}$ is the qubit relaxation rate, $\eta_\mathrm{f}=0.14$ is the total fluorescence measurement efficiency, $x(t)=\mathrm{Tr}(\sigma_x\rho_t)$ and $y(t)=\mathrm{Tr}(\sigma_y\rho_t)$ are the qubit Bloch coordinates corresponding to the density matrix $\rho_t$ (see Fig.~\ref{fig1}a) and $W_u$ and $W_v$ are two independent stochastic Wiener processes describing the measurement noise, which includes the zero point fluctuations, and such that $\dd W^2=\dd t$ and $\dd W$ is zero on average. Experimentally, the measurement takes a non infinitesimal time $\dd t=100~\mathrm{ns}$, which we chose smaller than the inverse measurement rates and compatible with the detection bandwidth (see Supplementary Note~4). 

Similarly, the dispersive detection setup (see Fig.~\ref{fig1}b) enables the measurement of a single quadrature $w(t)$ of the transmitted field at frequency $f_\mathrm{d}=f_\mathrm{r}-\chi_\mathrm{cq}/2$, which is between the cavity resonance frequencies $f_\mathrm{r}$ and $f_\mathrm{r}-\chi_\mathrm{cq}$ respectively corresponding to a qubit in the ground and excited state (the qubit and cavity are in the dispersive regime and $\chi_\mathrm{cq}= 5.1~\mathrm{MHz}$ as explained in Supplementary Note~2). The phase of the measured quadrature in the homodyne measurement can then be chosen in such a way that~\cite{campagneibarcq:tel-01248789,Weber2016}
\begin{equation}
w(t)\dd t=\sqrt{2\eta_\mathrm{d}\Gamma_{d}}z(t)\dd t+{\dd W_w(t)}.\label{wrecord}
\end{equation}
Importantly, the measurement induced dephasing rate $\Gamma_\mathrm{d}$ can be tuned arbitrarily as it is proportional to the drive power at $f_\mathrm{d}$. Similarly to the notations above, $\eta_\mathrm{d}=0.34$ is the total dispersive measurement efficiency, $z(t)=\mathrm{Tr}(\sigma_z\rho_t)$ is the last of the three Bloch coordinates (see Fig.~\ref{fig1}a) and $W_w$ is another independent Wiener process. 

\begin{figure}[!ht]
\includegraphics[width=9cm]{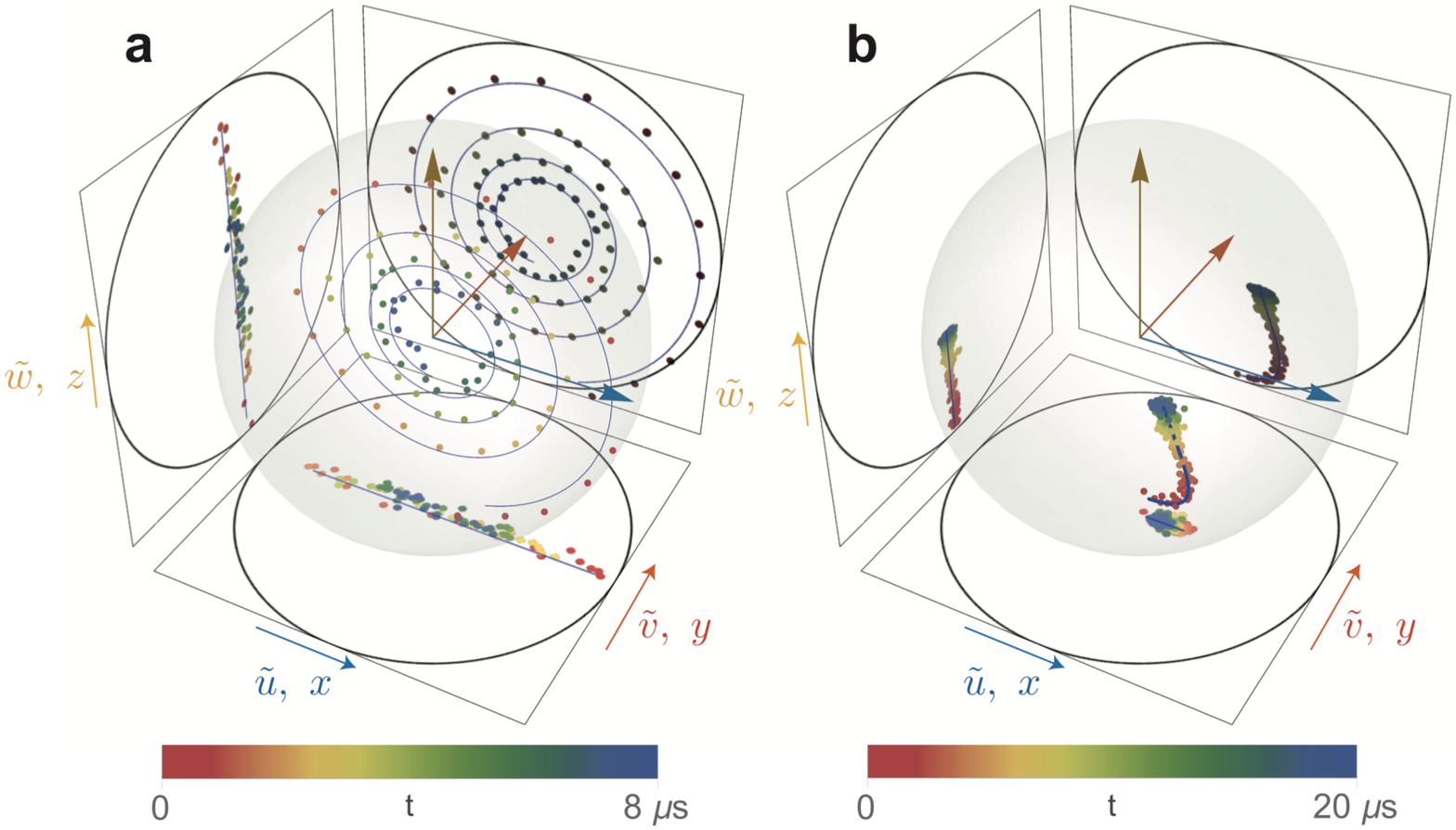}
\caption{Direct averaging of the three measurement records. \textbf{a} Dots: Rescaled average of the measurement records $\tilde{u}(t)=\bar{u}(t)/\sqrt{\eta_\mathrm{f}\Gamma_1/2}$, $\tilde{v}(t)=\bar{v}(t)/\sqrt{\eta_\mathrm{f}\Gamma_1/2}$ and $\tilde{w}(t)=\bar{w}(t)/\sqrt{2\eta_\mathrm{d}\Gamma_\mathrm{d}}$ for $1.5 ~10^6$ realizations of an experiment where the qubit starts in $|g\rangle$ at time $0$ and is driven so that it rotates at a Rabi frequency $\Omega/2\pi=(2~\mu\mathrm{s})^{-1}$ around $\sigma_y$ and endures a measurement induced dephasing rate $\Gamma_\mathrm{d}=(5~\mu\mathrm{s})^{-1}$. Lines: Calculated coordinates of the Bloch vector $x(t)$, $y(t)$ and $z(t)$ from the master equation (Eq.~\ref{SME} with $\eta_i=0$). \textbf{b} Same figure in the Zeno regime with a drive such that $\Omega/2\pi=(16~\mu\mathrm{s})^{-1}$ and $\Gamma_\mathrm{d}=(0.9~\mu\mathrm{s})^{-1}$.}
\label{average}
\end{figure}

\textbf{Full tomography by direct averaging.} As can be seen from Eqs.~(\ref{uvrecords},\ref{wrecord}), taking a raw average of the outcomes $(u,v,w)$ on a large number of realizations of the experiment directly leads to the Bloch coordinates $(x,y,z)$ of the qubit. In Fig.~\ref{average}, we show the direct averaging of the three outcomes in two configurations of the input drives: one in the regime of underdamped Rabi oscillations (Fig.~\ref{average}a) and another in the regime of strong dispersive measurement rate, the so-called Zeno regime (Fig.~\ref{average}b). The raw averaging of $(u,v,w)$, once rescaled by the prefactors in Eqs.~(\ref{uvrecords},\ref{wrecord}), agrees well with the average evolution of the qubit, as predicted by the solution of the master equation (Eq.~\ref{SME} below without the last stochastic term). We thus demonstrate that performing a dispersive measurement and a measurement of fluorescence reveals information on all three components of a spin-1/2. Such a direct full tomography cannot be done by measuring two records only~\cite{PhysRevX.6.011002,Hacohen-Gourgy2016}. Note however that it is possible to perform an indirect tomography using a small number of records and maximum likelihood estimation~\cite{Cook2014}. A comparison between our technique and the usual technique using a qubit rotation followed by a projective measurement is discussed in Supplementary Note~3.

\begin{figure}[!ht]
\includegraphics[width=9cm]{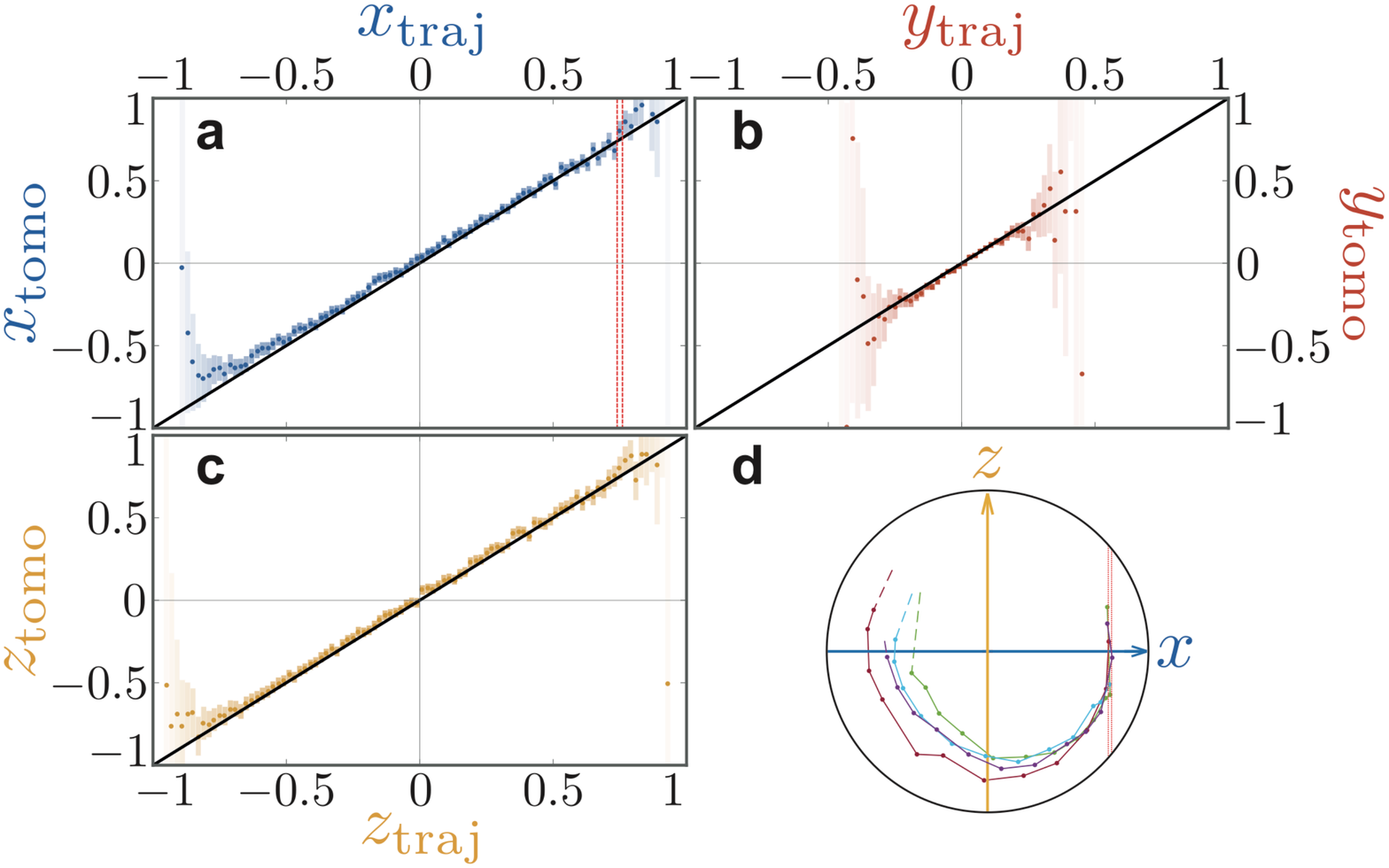}
\caption{Tomographic validation of the quantum trajectories. \textbf{a,b,c} Correlations between the coordinates $(x_\mathrm{traj},y_\mathrm{traj},z_\mathrm{traj})$ of the trajectories after $19.8\mathrm{~\mu s}$ of evolution and an independent tomography on the dataset corresponding to the experiment of Fig.~\ref{average}a. Each panel reprensents the average value of the tomography results for the subset of trajectories ending up less than 0.01 away from a given value of $x_\mathrm{traj}$ (\textbf{a}), $y_\mathrm{traj}$ (\textbf{b}) or $z_\mathrm{traj}$ (\textbf{c}). The error bars are given by the standard deviation of the tomography results divided by the square root of the number of trajectories in the subset (out of a total number of 1.5 million trajectories per panel). The agreement between the tomography and the coordinates of the trajectories demonstrates the validity of the quantum trajectories. \textbf{d} Bloch sphere representation of 3 quantum trajectories that end up with $0.74<x_\mathrm{traj}<0.76$ (red dashed line) after $19.8\mathrm{~\mu s}$ corresponding to one bin of the histogram in \textbf{a}.
}
\label{corr}
\end{figure}

Experiments of Figs.~\ref{average}a and \ref{average}b differ by the relative rate of the dispersive readout $\Gamma_\mathrm{d}$ compared to the Rabi frequency $\Omega$. For weak measurement rate $\Gamma_\mathrm{d},\Gamma_1<\Omega$, the Rabi oscillations are underdamped, while they are overdamped when $\Gamma_\mathrm{d}\gg\Omega,\Gamma_1$ owing to the fact that the Zeno effect prevents any unitary evolution such as Rabi oscillations. For a single realization of the experiment though, the trajectory of the qubit state that one can infer from the measurement records $u(t)$, $v(t)$ and $w(t)$ can strongly differ from this average behavior.

\textbf{Single quantum trajectories.} In order to determine this quantum trajectory, one can use the formalism of the stochastic master equation~\cite{campagneibarcq:tel-01248789}. The density matrix at time $t+\dd t$ can be decomposed into $\rho_{t+\dd t}=\rho_t+\dd\rho_t$, where
\begin{equation}
\dd\rho_t = i [\frac{\Omega}2\sigma_y,\rho_t]\dd t+\sum_k{{D}_k}(\rho_t)\dd t + \sum_k\sqrt{\eta_k} {\mathcal{M}_k}(\rho_t) {\dd W_k(t)}, 
\label{SME}
\end{equation}
with the four Lindblad superoperator ($k\in\{u,v,w,\varphi\}$)
\begin{equation}
{\mathcal{D}_k}(\rho_t) = L_k \rho_t L_k^\dagger - \frac{1}{2} \rho_t L_k^\dagger L_k - \frac{1}{2} L_k^\dagger L_k \rho_t,
\end{equation}
and the measurement backaction superoperator
\begin{equation}
{\mathcal{M}_k}(\rho_t) =  L_k \rho_t + \rho_t L_k^\dagger-\mathrm{Tr}(L_k \rho_t+\rho_t L_k^\dagger) \rho_t.
\end{equation}
In these expressions, the jump operators corresponding to heterodyning fluorescence are $L_u=\sqrt{\Gamma_1/2}\sigma_-$ and $L_v=i\sqrt{\Gamma_1/2}\sigma_-$ and the jump operator corresponding to homodyning the dispersive measurement is $L_w=\sqrt{\Gamma_\mathrm{d}/2}\sigma_z$. A fourth jump operator $L_\varphi=\sqrt{\Gamma_\varphi/2}\sigma_z$ corresponds to the unread ($\eta_\varphi=0$) pure dephasing of the qubit, so that the total decoherence rate $\Gamma_2=\frac{\Gamma_1}{2}+\Gamma_\varphi+\Gamma_\mathrm{d}$ can be tuned from $\Gamma_2=(11.2~\mu\mathrm{s})^{-1}$ to higher arbitrary values depending on the power of the drive at frequency $f_\mathrm{d}$. Interestingly, the two fluorescence measurement records $u$ and $v$ exert a different backaction but act identically on average (same Lindblad operators). The additional dispersive measurement that we introduced compared to Ref.~\cite{PhysRevX.6.011002} thus leads to a very different dynamics.

Using this formalism it is possible to reconstruct the quantum trajectory of the qubit state in time from any set of measurement records (see Fig.~\ref{fig1}a,c in the case where $\Omega/2\pi=(5.2~\mu\mathrm{s})^{-1}$ and $\Gamma_\mathrm{d}=(0.9~\mu\mathrm{s})^{-1}$). The validity of the reconstructed quantum trajectories can be tested independently by post-selecting an ensemble of realizations of the experiment for which the trajectory predicts a given value $x(T)=x_\mathrm{traj}$ at a time $T$. If the trajectories are valid, then a strong measurement of $\sigma_x$ at  time $T$ should give $x_\mathrm{traj}$ on average on this post-selected ensemble of realizations (Fig.~\ref{corr}a). We have checked for any value of $x_\mathrm{traj}$, $y_\mathrm{traj}$ and $z_\mathrm{traj}$ (Fig.~\ref{corr}), and for 30 representative configurations of drives that the trajectories predict the strong measurement results (Supplementary Note~1). In fact, we found that the agreement is verified for efficiencies $\eta_\mathrm{f}$ and $\eta_\mathrm{d}$ within a confidence interval of $\pm 0.02$ for any of the 30 configurations.

\begin{figure*}[!ht]
\includegraphics[width=18cm]{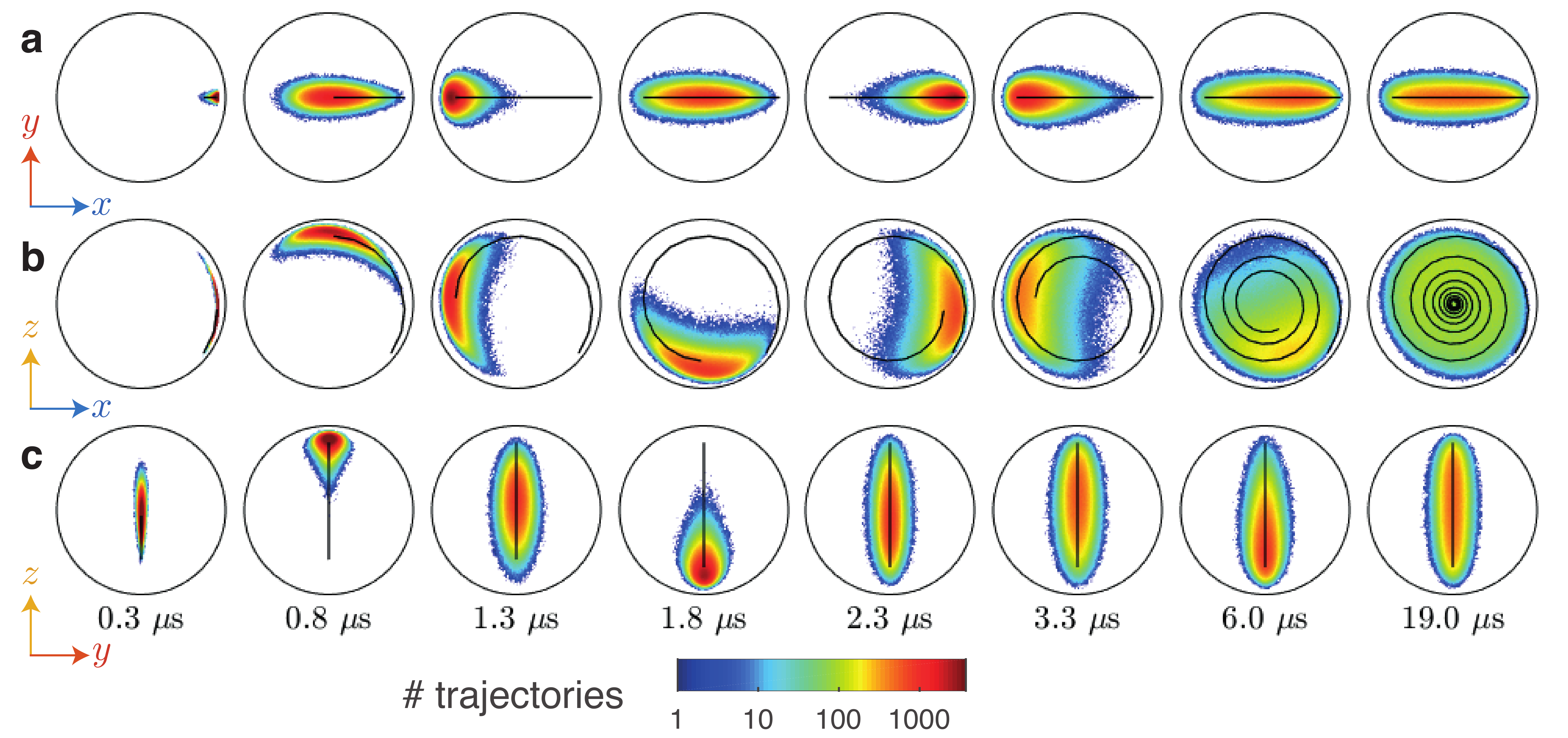}
\caption{Evolution of the distribution of quantum states. \textbf{a,b,c}  Colored dots: Each frame represents the marginal distribution, in the $x-y$ (\textbf{a}), $x-z$ (Fig.~\textbf{b}) and $y-z$ (Fig.~\textbf{c}) planes of the Bloch sphere, of the states of the qubit at a given time $\tau$ for 1.5 million realizations of the experiment, in the same experimental conditions as Fig.~\ref{average}a.
Each state $(x,y,z)$ is reconstructed from the measurement records $\{u(t),v(t),w(t)\}$ from time $t$ between 0 to $\tau$ using Eq.~(\ref{SME}). Time $\tau$ is increasing from $0.3~\mu\mathrm{s}$ to $19~\mu\mathrm{s}$ from left to right as indicated at the bottom of the figure. 
For each figure, the surrounding black circles represent the pure states of the plane (e.g. $z=0$ for \textbf{a}). Solid lines: average projection of all 1.5 millions of quantum trajectories $\{x(t), y(t), z(t)\}$ for $0.2~\mu\mathrm{s}<t<\tau$.
\label{figmovie}}
\end{figure*}

\textbf{Evolution of the distribution of states.} Any measurement record is a stochastic process and the corresponding quantum trajectories follow a random walk in the Bloch sphere with a state dependent diffusion constant. The inherent backaction of a quantum measurement is thus better discussed by representing distributions of states at a given time~\cite{Murch2013a,Hatridge2013,Tan2014,Weber2016,PhysRevX.6.011002,Naghiloo2016,Hacohen-Gourgy2016} or distributions of trajectories for a given duration~\cite{Weber2014,Jordan2015,Naghiloo2016a,Chantasri2016,Lewalle2017}.  Figure \ref{figmovie} gives a different perspective to the Rabi oscillation of Fig.~\ref{average}a by representing the distributions of the qubit states conditioned on the three measurement records $u(t)$, $v(t)$ and $w(t)$ for 1.5 million realizations of the experiment. In the Supplementary Note~1, one can find movies of the distributions of 1.5 millions  experimental realizations for each configuration of the Rabi frequency $\Omega$ and the dephasing rate $\Gamma_\mathrm{d}$ for a set of 30 different experimentally realized configurations. Evidently, the Rabi drive term $-\Omega\sigma_y/2$ still provides an overall angular velocity in the $x-z$ plane of the Bloch sphere. However, the measurement backaction is such that some trajectories are delayed while others are advanced compared to the average evolution. As time increases the spread in the qubit states grows as a result of the cumulated effect of the stochastic measurement backaction at each time step.

The effect of decoherence under a strong Rabi drive corresponds to an average loss of purity, defined as $\mathrm{Tr}(\rho^2)=(1+x^2+y^2+z^2)/2$ and it can be seen as a decreasing distance of the mean trajectory from the center of the Bloch sphere when time increases (solid line). When the dispersive measurement (dephasing channel) is measured in presence of the Rabi drive around $\sigma_y$, the corresponding distribution of states tends to be uniform in the $x-z$ plane at long times (right panel in Fig.~\ref{figmovie}b), which is similar to what is obtained by simultaneously measuring $\sigma_x$ and $\sigma_z$ in an effectively undriven qubit~\cite{Hacohen-Gourgy2016}. The experiment thus illustrates the fact that the average loss of purity  corresponds to the statistical uncertainty on the quantum state when the decoherence channel is unread.

\begin{figure}[!ht]
\includegraphics[width=9cm]{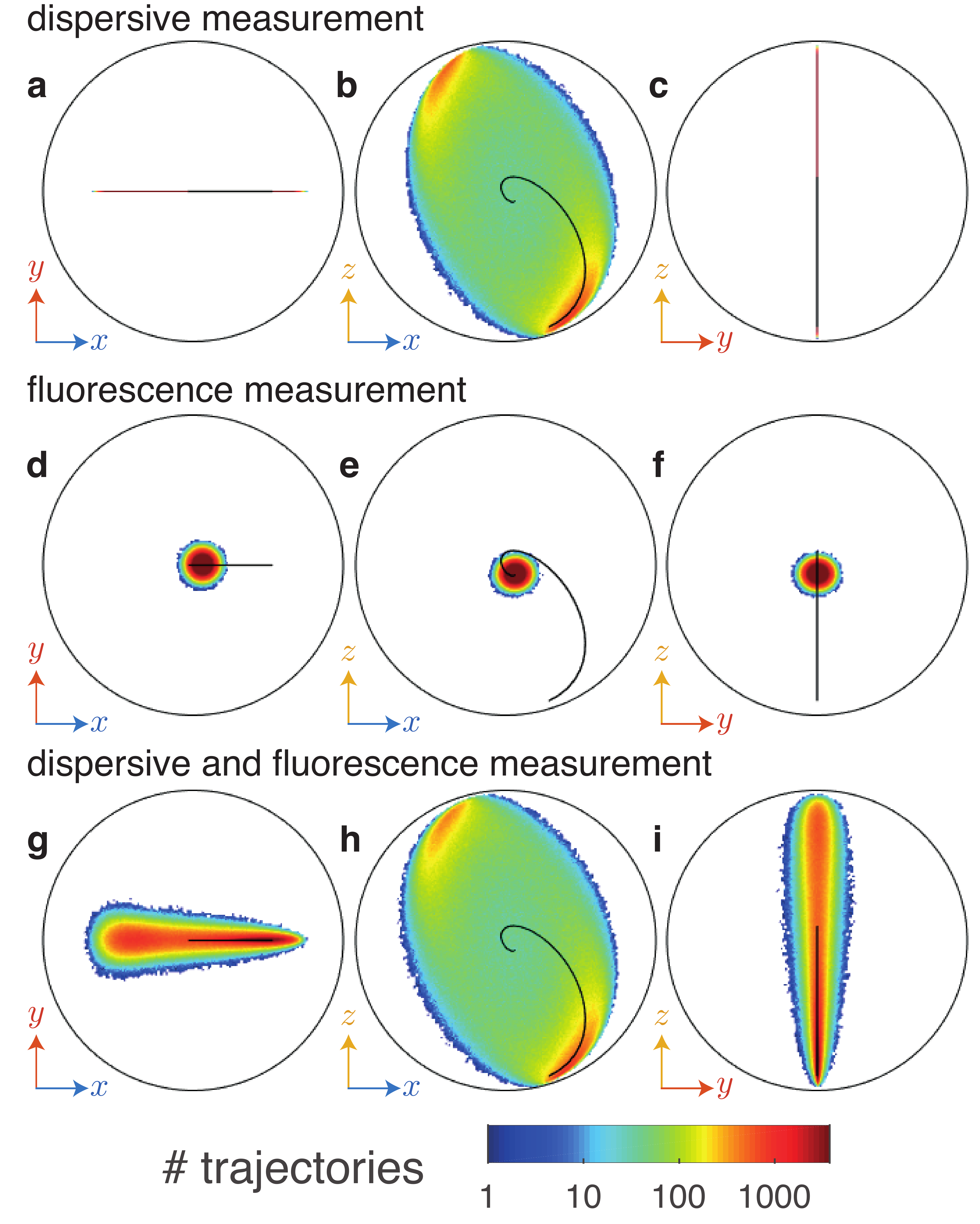}
\caption{Impact of the type of detector on the distribution of quantum states. \textbf{a,b,c} Marginal distribution in the $x, y$ (\textbf{a}), $x-z$ (\textbf{b}) and $y-z$ (\textbf{c}) planes of the Bloch sphere of the qubit states $\rho_\tau$ corresponding to 1.5 millions of measurement records at the cavity frequency only $\{w(t)\}$ from time $t$ between 0 and $\tau=6.5\mathrm{~\mu s}$. The information about $\{u(t),v(t)\}$ is here discarded ($\eta_\mathrm{f}=0$).  All panels in the figure correspond to the Zeno regime ($\Omega/2\pi=(5.2~\mu\mathrm{s})^{-1}\ll\Gamma_\mathrm{d}=(0.9~\mu\mathrm{s})^{-1}$)
As in Fig.~\ref{figmovie}, the boundary of the Bloch sphere is represented as a black circle and the average quantum trajectory as a solid line. \textbf{d,e,f} Case where the states are conditioned on fluorescence records $\{u(t),v(t)\}$ instead while discarding the information on $\{w(t)\}$ ($\eta_\mathrm{d}=0$). \textbf{g,h,i} Case where the states are conditioned on both fluorescence and dispersive measurement records $\{u(t),v(t),w(t)\}$. 
\label{figzeno}}
\end{figure}

\textbf{Interplay between detectors.} Interestingly, while the average trajectory stays in the $x-z$ plane with $\langle\sigma_y\rangle=0$, the backaction of the fluorescence measurement leads to a nonzero spread in the $y$ direction of the Bloch sphere. This competition between the backaction of relaxation (fluorescence measurement) and dephasing (dispersive measurement) measurements can be better observed when decoherence dominates the dynamics. In Fig.~\ref{figzeno}, we show the distributions of qubit states at a long time $\tau=6.5 \mathrm{~\mu s}$ after which the distribution is close to its steady state while the qubit is both Rabi driven and dispersively measured at a strong measurement rate. The trajectories are determined using three sets of measurement records: dispersive only $\{w(t)\}$, fluorescence only $\{u(t),v(t)\}$ or both. As in Fig.~\ref{average}b, the Zeno effect then leads to the dampening of the Rabi oscillations and the average trajectory (solid line) quickly reaches its steady state. 

In contrast, a trajectory corresponding to a single realization of the experiment where the dispersive measurement $w(t)$ is recorded is found to consist in a series of stochastic jumps between two areas of the Bloch sphere that are close to the two eigenstate of the $\sigma_Z$ measurement operator. In the distribution of states, this leads to two areas with high probability of occupation near the poles of the Bloch sphere. These areas can be interpreted as zones frozen by the Zeno effect. The rest of the Bloch sphere is still occupied with a lower probability (Fig.~\ref{figzeno}b) because of the finite time it takes for the jump to occur from one pole to the next under strong dispersive measurement rate~\cite{Choi2013}. Note how the ensemble of trajectories can go from uniform for weak measurement rates (Fig.~\ref{figmovie}b rightest panel) to localized at the poles for strong measurement rates (Fig.~\ref{figzeno}b).

As can be seen from Figs.~\ref{figzeno}a,c, the dispersive measurement alone does not provide any backaction towards the $y$ direction of the Bloch sphere so that the qubit states keeps a zero $\sigma_y$ component during its evolution. This is in stark contrast with the trajectories corresponding to measurement records $\{u(t),v(t)\}$ of the fluorescence (Figs.~\ref{figzeno}d-f), where at long times the qubit states spans a small ball in the Bloch sphere. Therefore, the combined action of Rabi drive and fluorescence measurement backaction leads to a uniform spread of the qubit state close to the most entropic state $\mathbf{1}/2$ at the center of the sphere. As expected, the quantum states that are conditioned on all measurement records $\{u(t),v(t),w(t)\}$ are less entropic than with a single measurement. This can be seen in Figs.~\ref{figzeno}g-i where the spread of the distributions is larger than for the cases of single measurements Figs.~\ref{figzeno}a-f. 

A clear asymmetry appears in the spread of the marginal distribution in the $x-y$ plane of Fig.~\ref{figzeno}g between positive and negative values of $x$. This asymmetry originates from the fact that the fluorescence measurement is linked to the jump operator $\sigma_-$ for which $|g\rangle$ is the single pointer state. Indeed the measurement backaction is null when the qubit state is close to $|g\rangle$ ($\mathcal{M}_u(|g\rangle\langle g|)=\mathcal{M}_v(|g\rangle\langle g|)=0$) while it is strongest when the qubit state is close to $|e\rangle$. Since the Rabi drive correlates the ground state to positive $x$ (red zone shifted to the right of the south pole in Fig.~\ref{figzeno}h) and the excited state to negative $x$, the spread in $y$ is smaller for positive $x$ than for negative $x$. This asymmetry highlights the profound difference between measuring both quadratures of fluorescence and measuring $\sigma_x$ and $\sigma_y$ simultaneously using dynamical states as in Ref.~\cite{Hacohen-Gourgy2016,Vool2016}. While both methods lead to the same result on average, their backaction differs. The latter corresponds to quantum non-demolition measurements, while fluorescence does not. In the end, the asymmetry in the distributions of Figs.~\ref{figzeno}g,i results from the incompatibility between a dispersive measurement with no backaction on $|e\rangle$ and a fluorescence measurement with maximal backaction on $|e\rangle$. 

In conclusion, we have shown quantum trajectories of a superconducting qubit reconstructed from three measurements originating from the simultaneous monitoring of its decoherence channels. It looks promising to test statistical properties of quantum trajectories~\cite{Franquet2017,Chantasri2017}, fluctuation relations in quantum thermodynamics~\cite{Gong2016,Elouard2017,Alonso2016,DiStefano2017,Deffner2016,Naghiloo2017,Liu2017}, quantum smoothing protocols~\cite{Tsang2009a,Tsang2009,Gammelmark2013a,Tan2014,Guevara2015,Tan2016,Tan2017}, and to perform parameter estimation~\cite{7403443,Kiilerich2016}.

\footnotesize

\textbf{Data availability}
The experiment was carried out for 30 experimental configurations with $\Omega/2 \pi$ ranging from $0$ to $(2 \mathrm{~\mu s})^{-1}$ and $\Gamma_d$ ranging from $(30 \mathrm{~\mu s})^{-1}$  to $(300 \mathrm{~ns})^{-1}$. All the experimental results can be visualized in a small animated application available online at

\url{http://www.physinfo.fr/publications/Ficheux1710.html}. 

The measurement can be chosen to take into account the measurement records of the dispersive measurement only, the fluorescence measurement only or both. The movies are also available to download at

\href{https://doi.org/10.6084/m9.figshare.6127958.v1}{https://doi.org/10.6084/m9.figshare.6127958.v1}. 

All raw data used in this study are available from the corresponding authors upon reasonable request.

\textbf{Author contributions}
Q.F. and S.J. contributed equally to this work. Q.F., S.J. and B.H. designed research and performed research; S.J. and Q.F. analyzed data; ZL contributed to the experimental setup; All authors wrote the paper.

\textbf{Acknowledgements}
We thank Philippe Campagne-Ibarcq, Michel Devoret, Andrew Jordan, Rapha\"el Lescanne, Mazyar Mirrahimi, Klaus M\o lmer, Pierre Rouchon, Alain Sarlette, Irfan Siddiqi and Pierre Six for fruitful interactions. Nanofabrication has been made within the consortium Salle Blanche Paris Centre. This work was supported by the EMERGENCES grant QUMOTEL of Ville de Paris.

\textbf{Competing Interests}
The Authors declare no Competing Financial or Non-Financial Interests.

\end{document}